\documentclass[10pt]{article} 
\usepackage[margin=1in]{geometry}

\usepackage[utf8]{inputenc} 
\usepackage{graphicx} 
\usepackage{amssymb}
\usepackage{verbatim} 
\usepackage {amsmath} 
\usepackage{bm} 

\usepackage{caption}
\usepackage{authblk}

\begin{document}

\title{Gravitational Hopfions}

\author[1]{Joe Swearngin}
\author[1,2]{Amy Thompson}
\author[2]{Alexander Wickes}
\author[1]{Jan Willem Dalhuisen}
\author[1,2]{Dirk Bouwmeester}

\affil[1]{\small Huygens Laboratory, Leiden University, PO Box 9504, 2300 RA Leiden, The Netherlands}
\affil[2]{\small Dept. of Physics, University of California, Santa Barbara, CA 93106}

\date{}

\maketitle

\begin{abstract}
The electromagnetic hopfion (EM hopfion) is a topologically nontrivial solution to the vacuum Maxwell equations with the property that any two field lines belonging to either the electric, magnetic, or Poynting vector fields (EBS fields) are closed and linked exactly once. The most striking and characteristic feature of this peculiar field configuration is the existence of an exceptional constant-time hyperplane wherein the EBS fields are tangent to three orthogonal Hopf fibrations. Using twistor methods we find the spin-$N$ generalization of the EM hopfion. Furthermore, we analyze the spin-2 solution within the framework of gravito-electromagnetism and show that the topology is manifest in the tendex and vortex lines. By decomposing the spin-2 field into spatial gravito-electric and gravito-magnetic tensors, we characterize its topological structure and evolution in terms of the EM hopfion.
\end{abstract}

\section{Introduction}
\label{intro}

Linked field configurations related to the Hopf fibration, called hopfions, have enjoyed successful application in many areas of physics including electromagnetism  \cite{Ranada2002,Irvine2008,Dalhuisen2012}\footnote{The term EM hopfion used here represents a departure from the standard nomenclature of ``EM knot'' used in  \cite{Ranada2002,Irvine2008,Dalhuisen2012}. However, as noted in \cite{Irvine2008} the field line structure of the EM hopfion is that of a collection of Hopf linked unknots and as such renders the term ''EM knot'' somewhat of a misnomer.}, magnetohydrodynamics \cite{Kamchatnov1982}, hadronic physics \cite{Skyrme1962,Faddeev1997}, helium superfluids \cite{Volovik1977}, and Bose-Einstein condensates \cite{Kawaguchi2008}. Among these hopfions the EM hopfion stands alone as the only radiative field. The aim of this work is to extend the notion of the EM hopfion to all spin-$N$ fields and in particular to spin-2 linearized gravitational fields. 

The salient feature of the Hopf structure of the EM hopfion is defined by the Poynting vector which is tangent to a Hopf fibration that propagates at the speed of light without deformation. The Poynting vector structure when extended to become a future pointing light-like 4-vector generates a set of shear-free space-filling null geodesics, collectively referred to as a Robinson congruence. In the twistor formalism the Robinson congruence features prominently as a framework to which many different physical fields are associated via the Penrose transform. From the twistor perspective \cite{PenroseSpinors2} all solutions of the massless spin-$N$ field equations in Minkowski space are seen to arise from the singularities of homogeneous functions on twistor space. Of particular note are the solutions known as \emph{elementary states}, fields which are constructed from homogeneous twistor functions whose vanishing defines a Robinson congruence \cite{Penrose1972}. The fields associated to these elementary states were introduced by Penrose, however this article is the first to establish the relation between the elementary states and the EM hopfion. Moreover, the relationship between the massless linear relativistic fields (massless Dirac, vacuum Maxwell, massless Rarita-Schwinger, and linearized vacuum Einstein equations), as expressed in the form of the massless spin-$N$ field equations, implies that all linear physical fields possess analogous topologically non-trivial field configurations.

We begin in section~\ref{sec:spinN} by presenting the spin-$N$ hopfion. In section~\ref{sec:spin1} we show that for $N=1$ this reproduces exactly the EM hopfion of Ref. \cite{Ranada2002}. In section~\ref{sec:spin2} the spin-2 solution is analyzed within the framework of gravito-electromagnetism  \cite{Maartens1998,Nichols2011}. We decompose the Weyl tensor into spatial gravito-electric and gravito-magnetic tensors and characterize their evolution in terms of the EM hopfion.  Since aspects of the twistor formalism, in particular the Penrose transform, will be essential tools for obtaining our results, we provide a brief overview of this material in the appendix.

\section{The Spin-$N$ Hopfion}
\label{sec:spinN}

As early as 1903 the importance of a complex analytic structure to the solutions of real PDE's began to emerge with the work of Whittaker, who found a contour integral expression for the general solution to Laplace's equation \cite{Whittaker1904}. In 1915 Bateman, building on the results of Whittaker, extended this analytic structure to obtain the general solution to the vacuum Maxwell equations \cite{Bateman1915}. In the twistor program of Penrose, the complex analytic structure is interpreted as encoding the geometry of spinor fields on spacetime \cite{Penrose1972}. In this language the various massless linear relativistic fields are represented as symmetric spinor fields (spin-$N$ fields) whose most general form is given by complex integral transforms of homogeneous twistor eigenfunctions of the helicity operator. The Penrose transform is a helicity-dependent complex integral transform which maps helicity eigenfunctions of homogeneity $-2h-2$ onto solutions of the massless spin-$h$ equation in Minkowski space,\footnote{We reserve $i$ and $j$ for spatial indices, other lower case Latin letters for Lorentz indices, upper case Latin letters and their primed variants for spinor and conjugate spinor indices respectively, and lower case Greek letters for twistor indices.}
\begin{equation}
\label{eqn:spinNfield}
\nabla^{AA'_1}\varphi_{A'_1\cdots A'_{2h}}(x) = 0.
\end{equation}
In this article we are concerned with real fields and so we can confine our discussion to the Penrose transform given by 
\begin{equation}
\varphi_{A'_1\cdots A'_{2h}}(x) = \frac{1}{2\pi i} \oint_\Gamma\pi_{A'_1}\cdots\pi_{A'_{2h}}f(Z)\pi_{B'}d\pi^{B'}
\end{equation}
and taken to be a contour integral on the Celestial sphere of light-like directions at $x$ \cite{PenroseSpinors2}. In the case of spin-1 and spin-2 the spinor fields $\varphi_{A'B'}(x)$ and $\varphi_{A'B'C'D'}(x)$  represent the Penrose transform of homogeneous twistor functions\footnote{Since the spinor fields $\varphi_{A'B'}(x)$ and $\varphi_{A'B'C'D'}(x)$ are obtained via complex contour integrals there exists a certain freedom in their twistor descriptions. A detailed exploration of this fact leads to the notion of complex sheaf cohomology.} of homogeneity -4 and -6 respectively and correspond to the $SL(2,\mathbb{C})$ representations of the field strength tensor $F_{ab}$ of electromagnetism and the Weyl tensor $C_{abcd}$ of general relativity. The simplest twistor functions corresponding to massless fields of helicity $h$ are called elementary states \cite{Penrose1972}, and the simplest non-trivial of these  is
\begin{equation}
\label{eqn:twistor_function}
f(Z) = (\overline{A}_\alpha Z^\alpha)^p (\overline{B}_\beta Z^\beta)^q,
\end{equation}
where $p,q < 0$ and $p+q = -2h-2$. Choosing $p = -1$ allows the pole due to the dual twistor $\overline{A}_{\alpha}$ to be simple and as we shall see later results in null EM fields and Petrov type N linearized gravitational fields. The details of the Penrose transform of equation (\ref{eqn:twistor_function}) are presented in the appendix and the result is 
\begin{equation}
\label{eqn:spinN_knot}
\varphi_{A'_1 \cdots A'_{2h}}(x) = \left( \frac{ 2 } {\Omega |x-y|^{2}} \right)^{2h+1}\mathcal{A}_{A'_{1}}\cdots\mathcal{A}_{A'_{2h}}
\end{equation}
where $\Omega$ is a constant scalar and $y$  a constant 4-vector each determined by the specific values of $\overline{A}_{\alpha}$ and $\overline{B}_{\beta}$, and $\mathcal{A}_{A'}$  is the spinor field associated to the Robinson congruence of  $A^{\alpha}$.

\section{The EM Hopfion}
\label{sec:spin1}

Classical electromagnetism resides in the spin-1 sector of equation (\ref{eqn:spinNfield}) \cite{PenroseSpinors1} with source-free field equation and field strength spinor given by
\begin{align}
\nabla^{AA'}\varphi_{A'B'} &= 0, \\
F_{A'B'AB} &= \varphi_{A'B'}\epsilon_{AB} + c.c. \label{eqn:EMspinorfield}
\end{align}
The standard electric and magnetic fields are recovered by decomposing $F_{ab}$ using a 4-velocity $u^a$ so that
\begin{align}
\mathcal{E}_b &= u^a F_{ab}, \\
\mathcal{B}_b &= -u^a \ast F_{ab}
\end{align}
where $\ast$ denotes the Hodge dual. Taking $u^a = (1,0,0,0)$ we have that
\begin{equation}
F_{ab} = %
	\begin{pmatrix}
	0 & E_x & E_y & E_z \\
	-E_x & 0 & -B_z & B_y \\
	-E_y & B_z & 0 & -B_x \\
	-E_z & -B_y & B_x & 0
	\end{pmatrix}.
\end{equation}
For $h=1$ equation (\ref{eqn:spinN_knot}) becomes
\begin{equation}
\varphi_{A'B'}(x) = \left( \frac{ 2 } {\Omega |x-y|^{2}} \right)^{3}\mathcal{A}_{A'}\mathcal{A}_{B'}
\end{equation}
where $\mathcal{A}_{A'}$ defines the doubly degenerate principle null direction of $F_{ab}$. Choosing
\begin{align}
\overline{A}_a &= (-i\sqrt{2},\sqrt{2},-i,1), \\
\overline{B}_\beta &= \frac{\pi^{1/3}}{2^{4/3}} (-\sqrt{2}, i\sqrt{2}, -1, i)
\end{align}
reproduces precisely the EM hopfion of Ra\~nada \cite{Ranada2002}. The expressions for the electric and magnetic fields can be conveniently expressed by a Riemann-Silberstein vector,
\begin{align}
\bm{F_R} &= \bm{E_R} + i \bm{B_R} \notag \\
	&= \frac{4}{\pi(-(t-i)^2+r^2)^3} %
	\begin{pmatrix}
	(x-iz)^2 - (t-i+y)^2 \\ 
	2(x-iz)(t-i+y) \\ 
	i(x-iz)^2 + i(t-i+y)^2 
	\end{pmatrix},
\end{align}
where $r^2=x^2+y^2+z^2$. The energy density and Poynting vector field are
\begin{align}
U_R &= \frac{16(1+x^2+(t+y)^2+z^2)^2}{\pi^2 (1 +2 (t^2+r^2) + (t^2 - r^2)^2)^3}, \label{eqn:EMenergy} \\
\bm{S_R} &=  \frac{U_R}{(1 + x^2 + (t+y)^2 + z^2)} %
	\begin{pmatrix} 
	2(x(t+y)+z) \\ 
	1 + (t+y)^2 - x^2 - z^2 \\ 
	2(z(t+y)-x) 
	\end{pmatrix}. \label{eqn:EMpoynting}
\end{align}
A visualization of these solutions is presented in row 1 of Figures \ref{fig:time0} and \ref{fig:time1} (see also \cite{Irvine2008}) and are here displayed for comparison with the GEM field configurations derived in the next section.

\section{The GEM Hopfion}
\label{sec:spin2}

Linearized general relativity makes up the spin-2 sector of the spin-$N$ equations \cite{PenroseSpinors1}. The field equation and associated field strength spinor are given by
\begin{align}
\nabla^{AA'}\varphi_{A'B'C'D'} &= 0, \\ 
C_{A'B'C'D'ABCD} &= \varphi_{A'B'C'D'}\epsilon_{AB}\epsilon_{CD} + c.c. \label{eqn:weyl_spinor}
\end{align}
In analogy with the spin-1 case, following Maartens \cite{Maartens1998} et al. and Nichols et al. \cite{Nichols2011}, we decompose the Weyl tensor using a 4-velocity $u^a$ into a tidal (gravito-electric) field and frame-drag (gravito-magnetic) field,
\begin{align}
\mathcal{E}_{ac} &= u^b u^d C_{abcd}, \\
\mathcal{B}_{ac} &= -u^b u^d \ast C_{abcd}. 
\end{align}
Taking $u^a = (1,0,0,0)$ we define the spatial tensors
\begin{align}
E_{ij} &= C_{i0j0} \label{eqn:spatial_tensors}, \\
B_{ij} &= - \ast C_{i0j0}. \notag
\end{align}
The symmetric traceless tensors $E_{ij}$ and $B_{ij}$ are called tidal and frame-drag fields since two orthogonal observers separated by a small spatial vector $\bm{\xi}$ will experience a relative tidal acceleration given by
\begin{equation}
\label{eqn:tidal_acceleration}
\Delta a_i = -E_{ij}\xi^j
\end{equation}
and a gyroscope at the tip of $\bm{\xi}$ will precess with angular velocity
\begin{equation}
\label{eqn:gyroscope_precession}
\Delta\Omega_i = B_{ij}\xi^j
\end{equation}
relative to inertial frames at the tail of $\bm{\xi}$. Since the tidal and frame-drag fields are symmetric and traceless, each may be characterized entirely by its eigenspaces. Thus, if $\bm{v}$ is an eigenvector of $E$ or $B$ then the integral curves of $\bm{v}$ are the gravitational analog of field lines. The tidal field stretches or compresses objects, and its associated field lines are referred to as tendex lines. The frame-drag field rotates gyroscopes, and its associated field lines are referred to as vortex lines. The tidal field has an associated eigenvalue $E_{\bm{v}}$ which has a physical interpretation given by the tidal acceleration equation (\ref{eqn:tidal_acceleration}). Thus, if the tendex eigenvalue is negative (respectively, positive) then an object oriented along the tendex line is stretched (compressed) along the tendex line. Similar relations hold for the frame-drag field whose eigenvalues are interpreted using equation (\ref{eqn:gyroscope_precession}), where an object oriented along a vortex line observes counter-clockwise (clockwise) precession of gyroscopes around the vortex line.

The gravito-electromagnetic hopfion (GEM hopfion) is constructed in the same fashion as the EM hopfion since linearized gravitational fields are taken to be spin-2 fields on $\mathbb{M}$. Thus, taking $h=2$ in equation (\ref{eqn:spinN_knot}), we have
\begin{equation}
\varphi_{A'B'C'D'}(x) = \left( \frac{ 2 } {\Omega |x-y|^{2}} \right)^{5}\mathcal{A}_{A'}\mathcal{A}_{B'}\mathcal{A}_{C'}\mathcal{A}_{D'}
\end{equation}
where $\mathcal{A}_{A'}$ defines the totally degenerate principle null directions of $C_{abcd}$ making it a Petrov type N linearized gravitational field. Taking the dual twistors $\overline{A}_\alpha$ and $\overline{B}_\beta$ to be the same as in the EM hopfion we construct the Weyl curvature from equation (\ref{eqn:weyl_spinor}). Then we perform the gravito-electric and gravito-magnetic decompositions as in equation (\ref{eqn:spatial_tensors}).

\begin{figure*}[t] 
\centering
\includegraphics{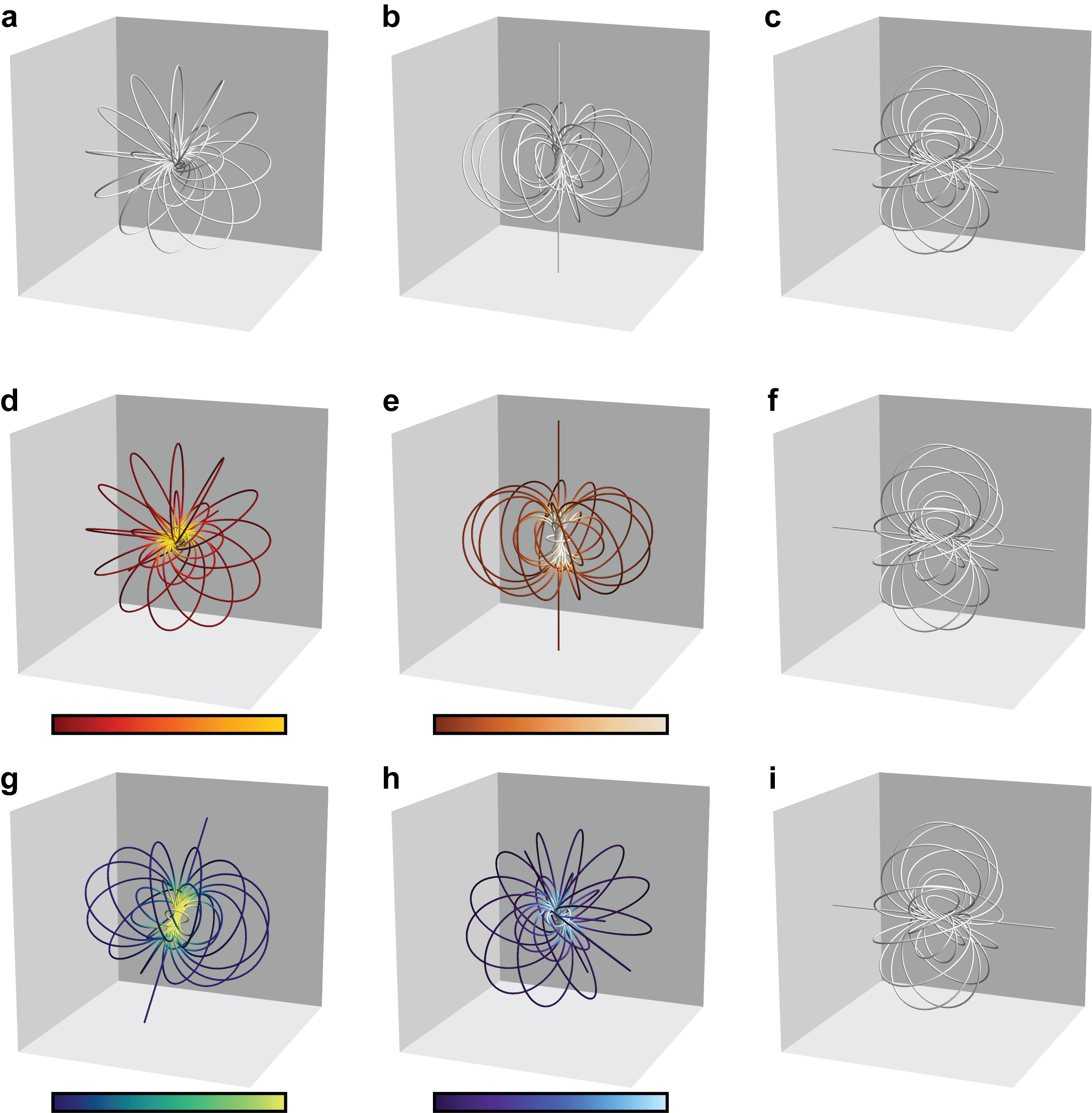}
\caption{A comparison of the spin-1 (EM) and spin-2 (gravity) hopfions at $t=0$. In the first row, we have the EM hopfion: \textbf{a} the electric field, \textbf{b} the magnetic field, and \textbf{c} the Poynting vector field. In the second row, we have the gravito-electric hopfion: \textbf{d} the negative eigenvalue field $E_-$, \textbf{e} the positive eigenvalue field $E_+$, and \textbf{f} the zero eigenvalue field $E_0$. In the third row, we have the gravito-magnetic hopfion: \textbf{g} the negative eigenvalue field $B_-$, \textbf{h} the positive eigenvalue field $B_+$, and \textbf{i} the zero eigenvalue field $B_0$. The color scale indicates magnitude of the eigenvalue, with lighter colors indicating a higher magnitude.}
\label{fig:time0}
\end{figure*}

 The tidal and frame-drag fields are characterized by their spectral decomposition which provides a three dimensional picture of the spacetime via the integral curves of their eigenvector fields and their physical interpretations given by equations (\ref{eqn:tidal_acceleration}) and (\ref{eqn:gyroscope_precession}). Performing the spectral decomposition we find that both fields possess an eigenvalue structure $\{+\Lambda,-\Lambda,0\}$ corresponding respectively to the eigenvectors $\{\bm{E_+},\bm{E_-},\bm{E_0}\}$ and $\{\bm{B_+},\bm{B_-},\bm{B_0}\}$, where
\begin{equation}
\Lambda(x) = \frac{2^{8/3} (1+x^2+(y+t)^2+z^2)^2}{\pi^{5/3}(1 +2 (t^2+r^2) + (t^2 - r^2)^2)^{5/2}}
\end{equation}
is the positive eigenvalue which determines the field strength, for both the tidal and frame-drag fields.

Considering first the eigenvector fields which correspond to the zero eigenvalue, we find that
\begin{align}
\bm{E_0} &= \bm{B_0} = %
	\begin{pmatrix} 
	2(x(t+y)+z) \\ 
	1+(t+y)^2-x^2-z^2 \\ 
	2(z(t+y)-x) 
	\end{pmatrix} \notag \\
	&= (1+x^2+(t+y)^2+z^2) \frac{\bm{S_R}}{|\bm{S_R}|}
\end{align}
which is, up to an overall scaling function, the Poynting vector of the EM hopfion. Constructing Riemann-Silberstein structures for the remaining fields
\begin{align}
\bm{F_{GE}} &= \bm{E_-} + i\bm{E_+}, \\
\bm{F_{GB}} &= \bm{B_-} + i\bm{B_+},
\end{align}
we find that
\begin{align}
\bm{F_{GE}} &= e^{i\pi/4} \bm{F_{GB}} \label{eqn:RS1} \\
	&= e^{i Arg(\vartheta)} \bm{F_R}, \label{eqn:RS2}
\end{align}
where
\begin{equation}
\vartheta = \sqrt{-(t-i)^2+r^2} \label{eqn:theta}.
\end{equation}

\begin{figure*}[t] 
\centering
\includegraphics{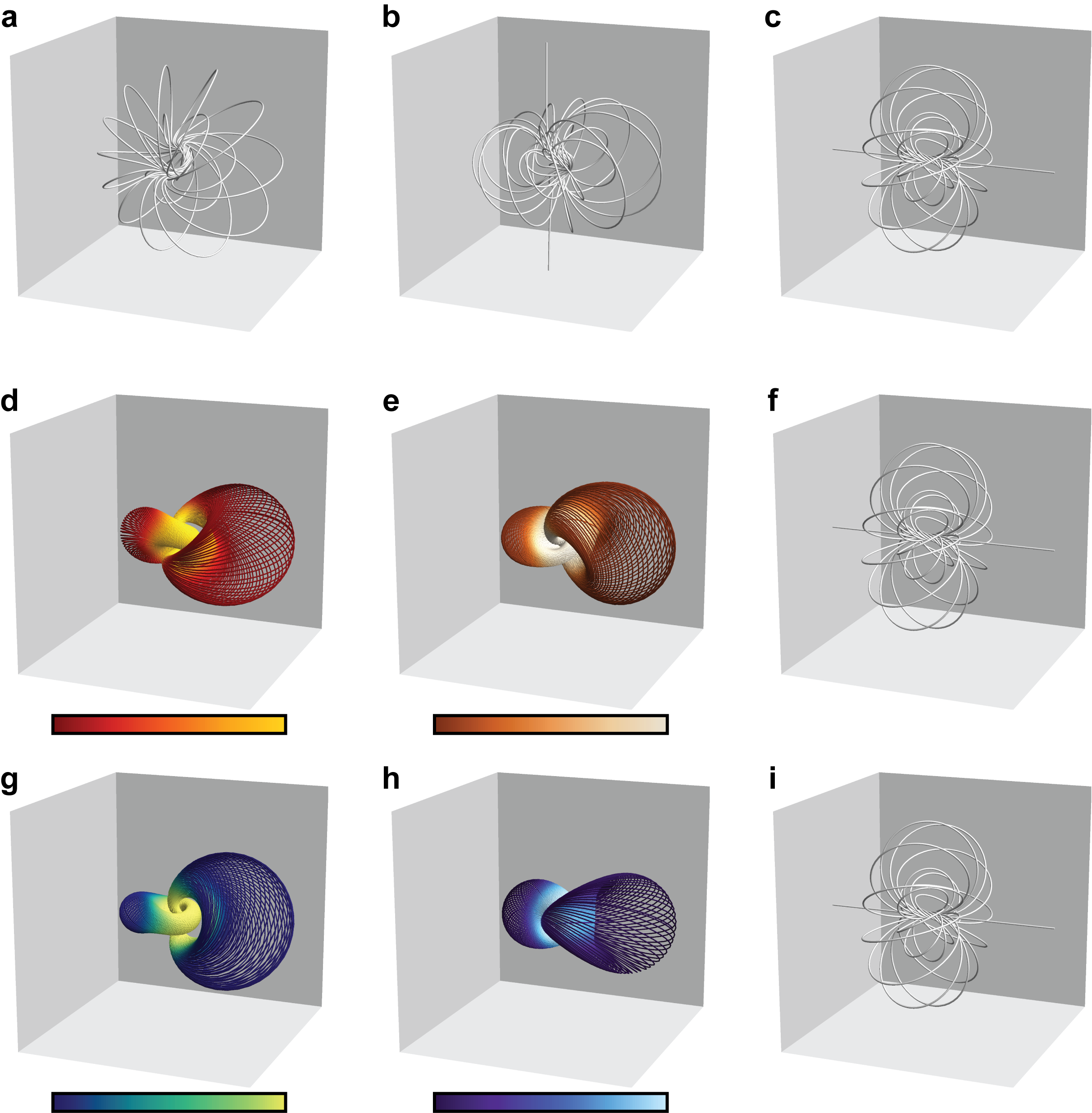}
\caption{A comparison of the spin-1 (EM) and spin-2 (gravity) hopfions at $t=1$, with layout the same as in Figure \ref{fig:time0}.}
\label{fig:time1}
\end{figure*}

This defines the tidal and frame-drag fields in terms of the EM hopfion. Equation (\ref{eqn:RS1}) shows that the frame-drag field is a rotation of the tidal field about the Poynting vector of the EM hopfion. Thus in passing to spin-2 we obtain two independent gravitational hopfion structures: a gravito-electric hopfion and a gravito-magnetic hopfion which differ simply by a global rotation. Equations (\ref{eqn:RS2}) and (\ref{eqn:theta}) show that the tidal field is a local duality transformation of the EM hopfion. At $t=0$, $Arg(\vartheta) =0$ and thus provides no duality transformation. Hence the tendex structure of the tidal field is the same as the field line structure of the EM hopfion. Furthermore, the vortex lines are the same as the tendex lines but rotated by $45^{\circ }$ in the $xy$-plane. Ergo, the $t=0$ tendex and vortex configuration is given by the six Hopf fibrations of Figure \ref{fig:time0} (rows 2 and 3) so that for each fixed eigenvalue the associated tendex (vortex) lines are closed and linked exactly once. For $t\neq 0$, $\vartheta $ is complex and hence the tendex lines differ from those of the EM hopfion by a local duality transformation as shown in Figure \ref{fig:time1} with the vortex structure different again by only a rotation of $45^\circ$.

For electromagnetic fields, there are two quantities that are invariant under local duality transformations: the energy density and Poynting vector. For the EM hopfion, these are given in equations (\ref{eqn:EMenergy}) and (\ref{eqn:EMpoynting}). In direct analogy with the local duality invariants of electromagnetism, a covariant super-energy density and super-Poynting vector that are invariant under a local duality transformation can be defined \cite{Maartens1998}. For the GEM hopfion, we find the super-energy density and super-Poynting vector are given by
\begin{align}
U_G &= \frac{1}{2}(E_{ab}E^{ab}+B_{ab}B^{ab}) \notag \\
	&= \frac{64\cdot 2^{1/3}(1+x^2+(t+y)^2+z^2)^4}{\pi^{10/3}(1 +2 (t^2+r^2) + (t^2 - r^2)^2)^5}, \\
( S_{G}) _a &= \epsilon_{abc}E_d^b B^{cd} \notag \\
	&= \frac{U_G}{(1+x^2+(t+y)^2+z^2)} %
	\begin{pmatrix} 
	2(x(t+y)+z) \\ 
	1+(t+y)^{2}-x^{2}-z^{2} \\ 
	2(z(t+y)-x) 
	\end{pmatrix}.
\end{align}
There is a striking similarity between the local duality invariants for the spin-1 and spin-2 cases, which differ only by the power of the scalar factor which shows the energy falls off more rapidly for the higher spin fields.

\section{Conclusion}
\label{conclusion}

According to Robinson, for each null shear-free geodesic congruence in Minkowski space there exists a null solution of Maxwell's equations \cite{Robinson1961}. Indeed, it has recently been shown using Kerr's theorem that the EM hopfion is derivable via the method of Robinson \cite{Dalhuisen2012}. Here we took a closer look at the relationship between the Robinson congruence of twistor theory and physical fields of spin-$N$. Using this construction, we showed that the EM hopfion represents the simplest non-trivial classical solution to the spin-1 massless field equation in twistor space. This enabled the extension of the spin-1 EM hopfion solution to the spin-$N$ equations. Taking the $N=2$ solution as a linearized Weyl tensor and aided by the concept of tendex and vortex lines, recently developed for the visualization of solutions in general relativity, we investigated the physical properties of the $N=2$ GEM hopfion and characterized its evolution in terms of the Riemann-Silberstein structure of the EM hopfion.

\section{Acknowledgements}

The authors would like to acknowledge discussions with Roger Penrose while he was the Lorentz Chair at the University of Leiden in the summer 2011. This work is supported by Marie Curie EXT-CT-042580 and NWO VICI 680-47-604.


\section*{Appendix A: 
Twistors and Total Momentum
}
\label{sec:twistors}

\begin{figure*}[t] 
\centering
\includegraphics{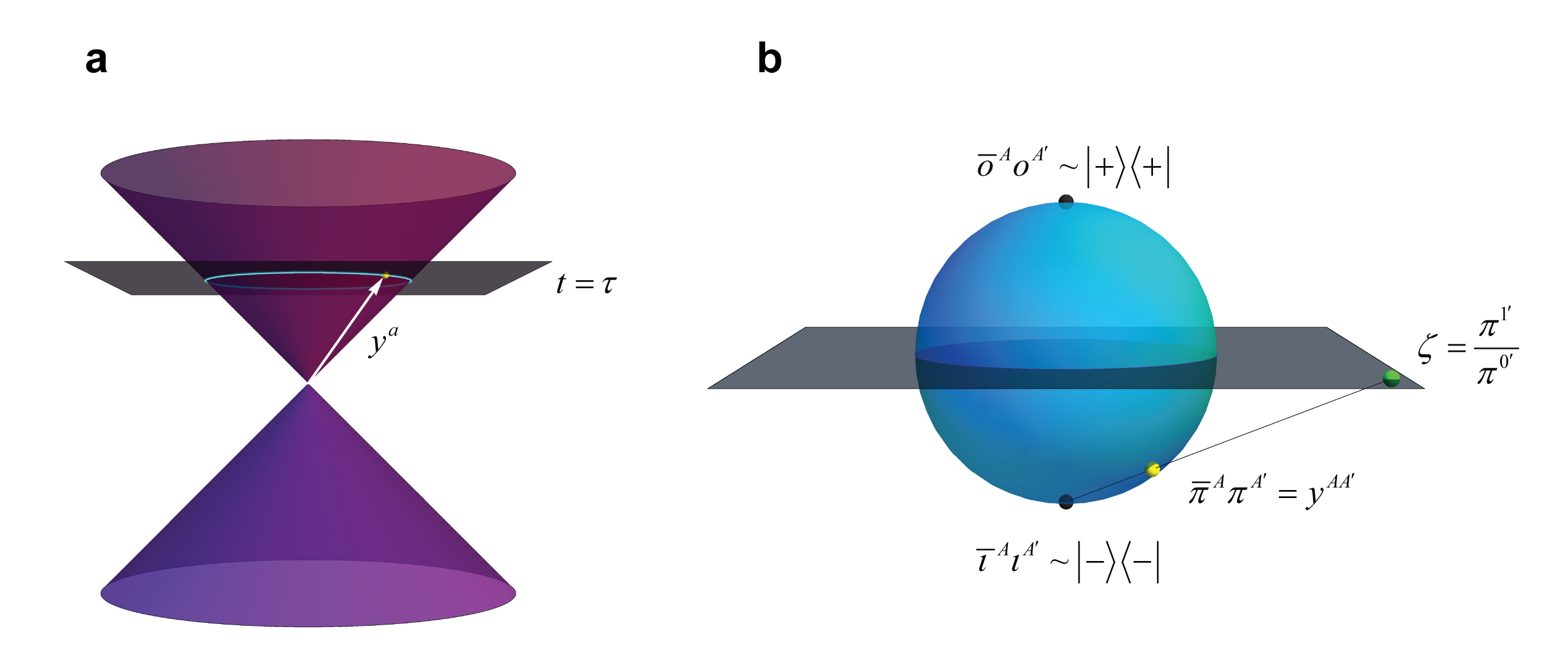}
\caption{The flagpole relation is the central relation in the spin geometry of $\mathbb{M}$. Here we see that to every 2-spinor $\pi^{A'}$ there corresponds a light-like 4-vector $y^a$ called its flagpole. \textbf{a} When the lightcone of the origin (violet) is intersected with the space-like hyperplane $t=\tau$ (dark grey), we obtain the Minkowski space representation of an expanding sphere of light emitted at the origin, called the celestial sphere (blue). Any light-like 4-vector $y^a$ (white) may be represented by the unique point where it meets this sphere (yellow). \textbf{b} Analogous to the pure state representation of the Bloch sphere, the celestial sphere may be put in 1-1 correspondence with the set of $2\times2$ Hermitian matrices with trace equal to $\sqrt{2} \tau$. The 2-spinor $\pi^{A'}$ is taken as a homogeneous coordinate on the complex plane (light grey) of the sphere $\zeta = \pi^{1'} / \pi^{0'}$ (green). The flagpole of $\pi^{A'}$ is then identified with the 4-vector $y^{AA'}$ written in flagpole form as a $2\times2$ Hermitian matrix.}
\label{fig:bloch}
\end{figure*}

The fundamental relationship between relativistic $SL(2,\mathbb{C})$ 2-spinors and light-like 4-vectors, as illustrated in Figure \ref{fig:bloch}, defines the spin geometry of Minkowski space $\mathbb{M}$ and underlies the entire twistor program laid out by Penrose \cite{Penrose1967}.

One can think of twistor space $\mathbb{T}$ as the total momentum space for massless particles \cite{PenroseSpinors1} in which the linear and angular momenta are combined into a single object called a twistor. This is accomplished by using their $SL(2,\mathbb{C})$ representations. It turns out that, combined, linear and angular momentum possess two spinorial degrees of freedom denoted by $\omega ^{A}$ and $\pi _{A^{\prime }}$. Since the linear momentum $p^{a}$ is light-like it can be decomposed as the flagpole $\overline{\pi }^{A}\pi ^{A^{\prime }}$ of some spinor $\pi _{A^{\prime }}$. The second spinor $\omega^{A}$ is related to the relativistic angular momentum $M^{ab}$ in that $i\omega^{(A}\overline{\pi }^{B)}$ and its conjugate represent the self-dual (SD) and anti-self-dual (ASD) parts of $M^{ab}$ respectively (where the parentheses denote normalized symmetrization over the indices). The dependence of the momentum structure $(M^{ab},p^{c})$ on the origin induces a position dependence in $(\omega ^{A},\pi _{A^{\prime }})$, making them spinor fields on spacetime defined by
\begin{align}
\pi_{A'}(x) &= \pi_{A'}, \\
\omega^A(x) &= \omega^A - ix^{AA'}\pi_{A'},
\end{align}
where $x^{AA^{\prime }}$ is the $SL(2,\mathbb{C})$ representation of the 4-vector $x^{a}=(t,x,y,z)$ given by 
\begin{equation}
x^{AA'} = \frac{1}{\sqrt{2}} %
	\begin{pmatrix}
	t+z & x+iy \\
	x-iy & t-z
	\end{pmatrix}.
\end{equation}
The $SL(2,\mathbb{C})$ representation of a 4-vector allows an algebraic realization of the flagpole relation of Figure \ref{fig:bloch} where we associate with the spinor $\pi _{A^{\prime }}$ a light-like 4-vector $y^{a}=(y^{0},y^{1},y^{2},y^{3})$, called its flagpole, which has coordinates defined by 
\begin{equation}
\overline{\pi}^A\pi^{A'} = \frac{1}{\sqrt{2}} %
	\begin{pmatrix}
	y^0+y^3 & y^1+iy^2 \\
	y^1-iy^2 & y^0-y^3
	\end{pmatrix}.
\end{equation}

We now define a twistor $Z^\alpha$ as the momentum structure $ (M^{ab},p^c)$ encoded in a kinematically related pair of 2-spinors $(\omega^A,\pi_{A^{\prime }})$ such that
\begin{enumerate}
\item the flagpole of $\pi_{A'}$ represents a linear momentum
\begin{equation}
\label{eqn:linear_momentum_correspondence}
p^{AA'} = \overline{\pi}^A\pi^{A'},
\end{equation}
and
\item the symmetrized product $i\omega^{(A}\overline{\pi}^{B)}$ and its conjugate represent respectively the SD and ASD components of the angular momentum tensor,
\begin{equation}
\label{eqn:angular_momentum_correspondence}
M^{AA' BB'} = i\omega^{(A}\overline{\pi}^{B)}\epsilon^{A'B'} + c.c.
\end{equation}
\end{enumerate}

The kinematic pair $(\omega^A,\pi_{A'})$ comes together to yield a 4-dimensional complex object, called a twistor, with components
\begin{align}
(Z^0,Z^1) &\equiv (\omega^0,\omega^1), \\
(Z^2,Z^3) &\equiv (\pi_{0'},\pi_{1'}), 
\end{align}
which is commonly abbreviated $Z^\alpha = (\omega^A,\pi_{A'})$. The twistor dual is given by $\overline{Z}_\alpha = (\overline{\pi}_A,\overline{\omega}^{A'})$ and allows one to construct a very useful expression for the helicity associated to the momentum structure $Z^\alpha$ given by
\begin{equation}
\label{eqn:twistor_helicity}
\mathfrak{h} = \frac{1}{2}Z^\alpha\overline{Z}_\alpha.
\end{equation}
Helicity provides for a natural geometric partition of $\mathbb{T}$ into three parts, $\mathbb{T}^+$, $\mathbb{T}^-$, and $\mathbb{N}$, according to whether the helicity is positive, negative, or zero. Twistors from $\mathbb{N}$ are said to be null, otherwise they are non-null. Only null twistors have a direct geometric correspondence with $\mathbb{M}$.

\begin{figure*}[t] 
\centering
\includegraphics{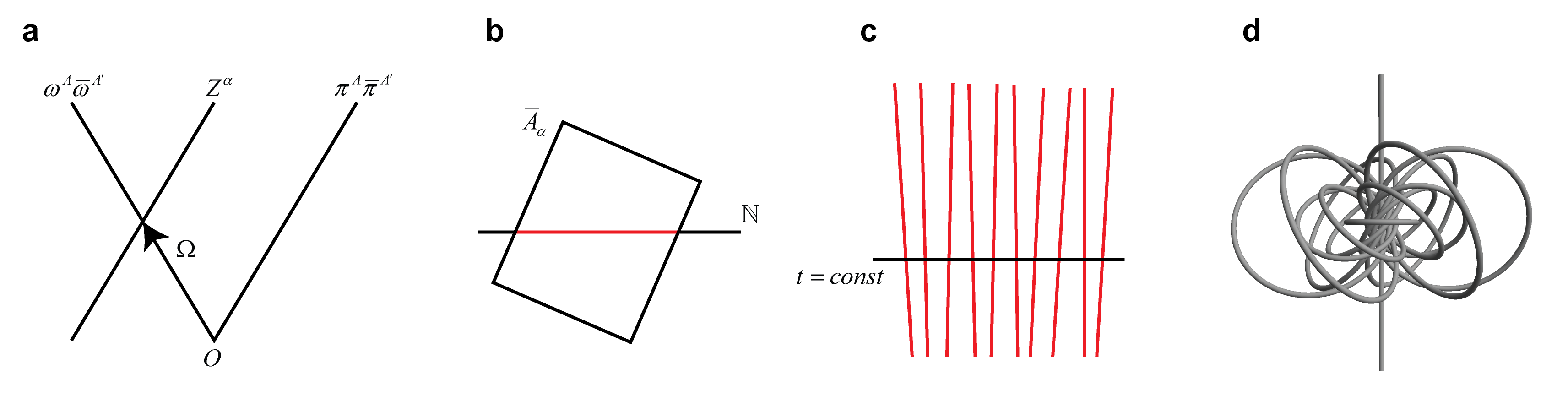}
\caption{A null twistor $Z^\alpha$ corresponds to a light-like world-line in $\mathbb{M}$. A non-null twistor $A^\alpha$ possesses only an indirect correspondence in $\mathbb{M}$ constructed via the direct correspondence for null twistors. \textbf{a} If $Z^\alpha = (\omega^A, \pi_{A'})$ is a null twistor, then it corresponds to a light-like world-line (null geodesic) in $\mathbb{M}$, parallel to the flagpole of $\pi_{A'}$, which meets the lightcone of the origin along the light-like ray parallel to the flagpole of $\omega^A$. Here $\Omega = x^{AA'}(0)$. \textbf{b} The dual of a non-null twistor $\overline{A}_\alpha$ is a plane in $\mathbb{T}$ which is uniquely defined by its intersection with $\mathbb{N}$ (red). \textbf{c} The Minkowski space representation of the non-null twistor $\overline{A}_\alpha$ wherein each of the null twistors in the intersection with $\mathbb{N}$ appear as the null geodesics in $\mathbb{M}$ which comprise the Robinson congruence. \textbf{d} Viewed on a hyperplane of constant time, the Robinson congruence defines a vector field whose integral curves are the fibers of a Hopf map projected down stereographically onto the hyperplane.}
\label{fig:twistorsMinkowski}
\end{figure*}

Twistor geometry in $\mathbb{M}$ is determined by the vanishing of $\omega^A(x)$, yielding the incidence relation 
\begin{equation}
\label{eqn:incidence_relation}
\omega^A = ix^{AA'}\pi_{A'}.
\end{equation}
This relation can be solved for Hermitian $x^{AA^{\prime }}$, and thus points in $\mathbb{M}$, only when $Z^\alpha$ is null. In that case the solution is 
\begin{equation}
\label{eqn:incidence_solution}
x^{AA'}(r) = \frac{\omega^A\overline{\omega}^{A'}}{i\overline{\omega}^{B'}\pi_{B'}} + r\overline{\pi}^A\pi^{A'}
\end{equation}
for arbitrary real $r$. The geometry of equation (\ref{eqn:incidence_solution}) is shown in Figure \ref{fig:twistorsMinkowski} and represents a light-like worldline in $\mathbb{M}$ parallel to the flagpole of $\pi_{A^{\prime }}$, which meets the lightcone of the origin along the flagpole of $\omega^A$ at the point $x^{AA^{\prime }}(0)$.

For non-null twistors, equation (\ref{eqn:incidence_relation}) possesses no Hermitian solutions, and thus its direct geometric interpretation is in terms of complexified Minkowski space. However, as shown in Figure \ref{fig:twistorsMinkowski}, we can obtain an indirect geometric correspondence by exploiting the more direct relation for null twistors. If $A^\alpha$ is non-null then its dual $\overline{A}_\alpha$ is a plane in $\mathbb{T}$. This plane is uniquely defined by its intersection with $\mathbb{N}$ and hence by the solution of the system
\begin{align}
\overline{A}_\alpha Z^\alpha &= 0, \\
\overline{Z}_\alpha Z^\alpha &= 0. 
\end{align}
By equation (\ref{eqn:incidence_solution}), each point in this intersection represents a null geodesic. The set of all null twistors in the intersection corresponds to a space-filling set of null geodesics in $\mathbb{M}$ called the Robinson congruence. If we project the tangent vector field of one of these congruences onto a hyperplane of constant $t$ then we find that it forms the tangent vector field of a Hopf fibration. The integral curves of the vector field are a special family of circles which lie on a set of space-filling nested tori. Each circle, called a Villarceau circle, is linked with every other one exactly once, and the entire structure propagates with the speed of light along its axis of symmetry without deformation. In fact, it was the Hopf structure of the Robinson congruence which inspired Penrose with the name ``twistor.''

\begin{figure}[t] 
\centering
\includegraphics{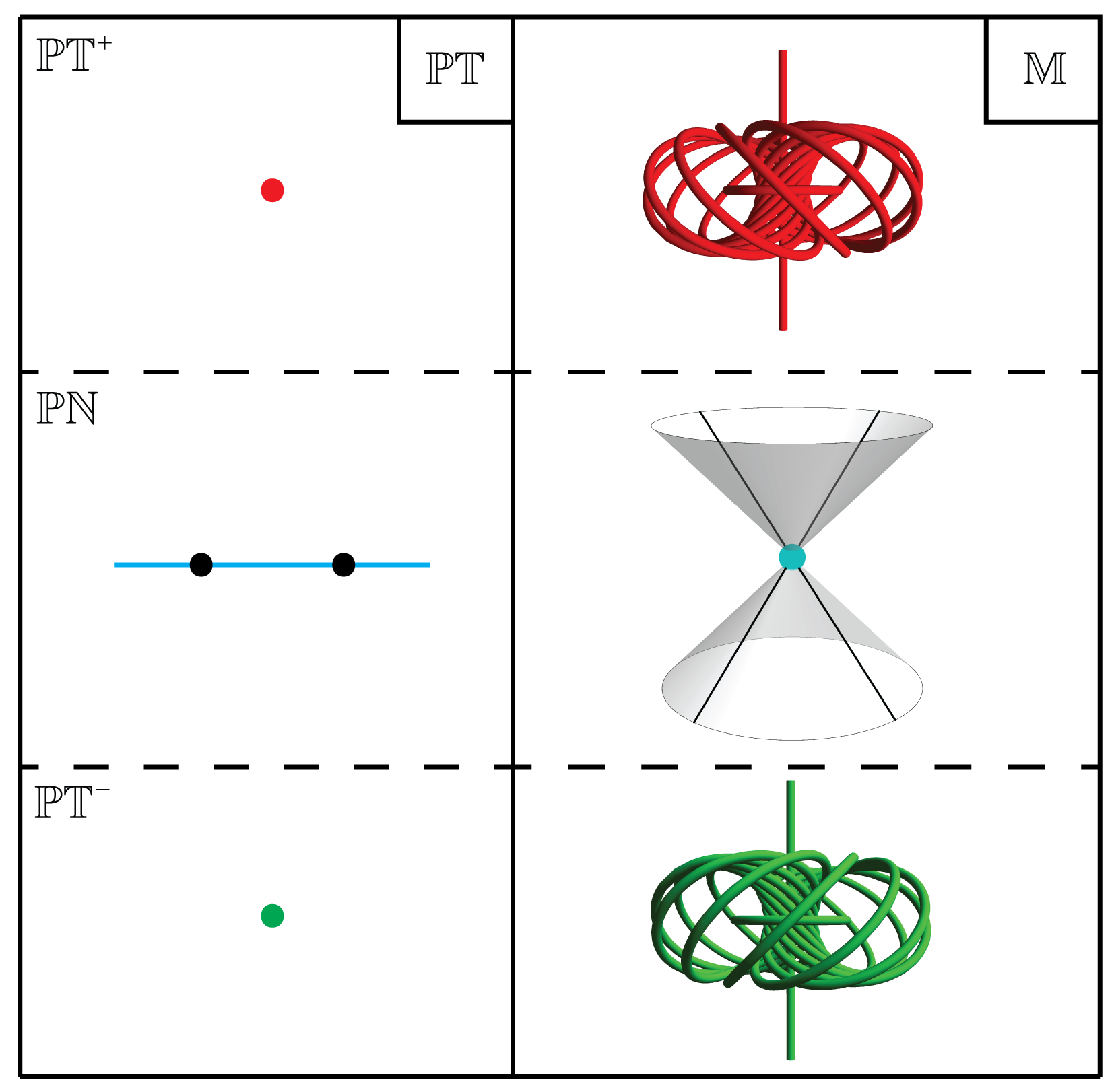}
\caption{A summary of the geometric correspondences in terms of $\mathbb{PT}$ and $\mathbb{M}$. A point in either $\mathbb{PT}^+$ (red) or $\mathbb{PT}^-$ (green) corresponds to a Robinson congruence in $\mathbb{M}$. An entire $\mathbb{CP}^1$ (celestial sphere) in $\mathbb{PN}$ (blue) corresponds to a point in $\mathbb{M}$. Any two points which lie on a $\mathbb{CP}^1$ in $\mathbb{PN}$ correspond to light-like world-lines which lie on the lightcone of the point corresponding to that $\mathbb{CP}^1$.}
\label{fig:projectiveTwistors}
\end{figure}

The momentum correspondences, equations (\ref{eqn:linear_momentum_correspondence}) and (\ref{eqn:angular_momentum_correspondence}), clearly possess $U(1)$ invariance. Yet the geometric correspondence defined by the incidence relation, equation (\ref{eqn:incidence_relation}), is projective. That is, a twistor $Z^\alpha$ multiplied by any non-zero complex number corresponds to the same object in $\mathbb{M}$ as $Z^\alpha$ itself. This leads us to consider as most fundamental the projective twistor space $\mathbb{PT}$ defined as $\mathbb{T}$ under the equivalence relation $Z^\alpha \sim \lambda Z^\alpha$ for $\lambda \neq 0$. The partitions of $\mathbb{T}$ naturally extend to partitions of $\mathbb{PT}$, denoted by $\mathbb{PT}^+$, $\mathbb{PT}^-$, and $\mathbb{PN}$. The geometric correspondences are most succinctly characterized in terms of the partitions of $\mathbb{PT}$, as shown in Figure \ref{fig:projectiveTwistors}.

\section*{Appendix B: 
Twistor Quantization and Holomorphic Helicity Eigenfunctions
}
\label{sec:quantization}

The approach to twistor quantization proceeds in much the same way as in standard quantum theory. We first identify a set of canonical variables. Next we impose commutation relations on these variables. Once commutation relations are imposed, we fix a representation in which half of the variables become derivatives and the other half remain parameters. Fixing this representation allows us to construct explicit equations for some observable of interest for which we can seek eigenfunctions parameterized by only one of the canonical variables.

Choosing $Z^\alpha$ and $\overline{Z}_\alpha$ as our canonical variables, we impose the commutation relations
\begin{align}
[Z^\alpha,Z^\beta] &= [\overline{Z}_\alpha,\overline{Z}_\beta] = 0, \\
[Z^\alpha,\overline{Z}_\beta] &= \delta^\alpha_\beta.
\end{align}
While this may seem rather ad hoc, it turns out that these relations, under the momentum correspondences, equations (\ref{eqn:linear_momentum_correspondence}) and (\ref{eqn:angular_momentum_correspondence}), induce the standard commutation relations between $M^{ab}$ and $p^c$ which generate the Poincar\'e group. Therefore, twistor quantization can be consistently viewed as total momentum quantization.

Next, we are faced with a choice of representation
\begin{equation}
\label{eqn:choiceZ}
(Z^\alpha,\overline{Z}_\beta) = (Z^\alpha,-\frac{\partial}{\partial Z^\beta})
\end{equation}
or
\begin{equation}
(Z^\alpha,\overline{Z}_\beta) = (\frac{\partial}{\partial \overline{Z}_\alpha},\overline{Z}_\beta).
\end{equation}
Once a representation is chosen, the eigenfunctions are confined to be functions of only one of $Z^\alpha$ or $\overline{Z}_\beta$. This is tantamount to specifying that the eigenfunctions be either holomorphic,
\begin{equation}
\label{eqn:holomorphic}
\frac{\partial}{\partial \overline{Z}_\beta}f = 0,
\end{equation}
or antiholomorphic,
\begin{equation}
\frac{\partial}{\partial Z^\alpha}f = 0,
\end{equation}
in the twistor parameter $Z^\alpha$. Following Penrose \cite{Penrose1972} we choose the holomorphic case defined by equations (\ref{eqn:choiceZ}) and (\ref{eqn:holomorphic}).

For massless free fields, helicity is the only quantum number and so the dynamics is entirely determined by the eigenfunctions of the helicity equation (\ref{eqn:twistor_helicity}). Polarizing to avoid factor ordering issues and converting to our chosen representation, we have that
\begin{align}
\label{eqn:holomorphic_helicity}
\mathfrak{h} &= \frac{1}{4}(Z^\alpha\overline{Z}_\alpha + \overline{Z}_\alpha Z^\alpha) \notag \\
	&= -\frac{1}{2}(Z^\alpha\frac{\partial}{\partial Z^\alpha} + 2). 
\end{align}
The operator $z\frac{\partial}{\partial z}$ is known as the Euler homogeneity operator and its eigenfunctions satisfy
\begin{align}
z\frac{\partial}{\partial z}f(z) = n f(z), \\
f(tz) = t^n f(z).
\end{align}
We recognize equation (\ref{eqn:holomorphic_helicity}) as representing a shifted Euler homogeneity operator on twistor space, and thus its eigenfunctions are homogeneous twistor functions where the relation between the homogeneity $n$ and the eigenvalue $h$ is given by
\begin{equation}
\label{eqn:homogeneity}
n = -2h - 2.
\end{equation}

\section*{Appendix C: 
The Penrose Transform
}
\label{sec:transform}

The Penrose transform is a helicity-dependent integral transform which maps the holomorphic eigenfunctions of the helicity operator onto solutions of the spin-$N$ field equations on $\mathbb{M}$ \cite{Penrose1972}. Since we are interested only in real fields of spin-1 and spin-2, and thus fields whose SD and ASD components are conjugate, we require only the Penrose transform for positive helicity,
\begin{equation}
\label{eqn:penroseTransform}
\varphi_{A'_1\cdots A'_{2h}}(x) = \frac{1}{2\pi i} \oint_\Gamma\pi_{A'_1}\cdots\pi_{A'_{2h}}f(Z)\pi_{B'}d\pi^{B'}
\end{equation}
where $\Gamma$ is a contour on the Celestial sphere of $x$ (Figure \ref{fig:bloch}b) which separates the poles of $f(Z)$. The result is a spinor field which satisfies the spin-$N$ massless field equation \cite{PenroseSpinors1}
\begin{equation}
\nabla^{AA'_1}\varphi_{A'_1\cdots A'_{2h}}(x) = 0.
\end{equation}

Since the helicity operator is a shifted Euler homogeneity operator on $\mathbb{T}$, we have that the positive helicity eigenfunctions of equation (\ref{eqn:holomorphic_helicity}) are twistor functions of negative homogeneity. The Penrose transform is manifestly a contour integral on the celestial sphere of $x$ so there must be at least two poles in order for the integral to be non-vanishing. The simplest function which is homogeneous of degree $-2h-2$ with two distinct poles is
\begin{equation}
f(Z) = (\overline{A}_\alpha Z^\alpha)^p (\overline{B}_\beta Z^\beta)^q,
\end{equation}
where $p,q<0$ and $p+q=-2h-2$.

Consider the Penrose transform with $f(Z)$ given by
\begin{equation}
f(Z) = (\overline{A}_\alpha Z^\alpha)^{-1} (\overline{B}_\beta Z^\beta)^{-2h-1}.
\end{equation}
Let $\overline{A}_\alpha = (\mu_A,\lambda^{A'})$, $\overline{B}_\alpha = (\sigma_A,\psi^{A'})$ be dual twistors such that
\begin{align}
\overline{A}_\alpha Z^\alpha &= i\mu_A x^{AA'} \pi_{A'} + \lambda^{A'} \pi_{A'}  \notag \\
	&\equiv \mathcal{A}^{A'} \pi_{A'}, \\
\overline{B}_\beta Z^\beta &= i\sigma_B x^{BB'} \pi_{B'} + \psi^{B'} \pi_{B'}  \notag \\
	&\equiv \mathcal{B}^{B'} \pi_{B'},
\end{align}
where $Z^\alpha = (ix^{AA'} \pi_{A'}, \pi_{A'})$. Now
\begin{align}
\pi_{C'} d\pi^{C'} &= \pi_{C'} d\pi_{D'} \epsilon_{D'C'}  \notag \\
	&= \pi_{0'} d\pi_{1'} - \pi_{1'} d\pi_{0'}  \notag \\
	&= (\pi_{0'})^2 d(\frac{\pi_{1'}}{\pi_{0'}}).
\end{align}
Introducing the canonical spin bases $\{o_{A'}, \iota_{A'}\}$ into the primed spin space $S'$ we have that
\begin{align}
\pi_{A'} &= \pi_{0'} o_{A'} + \pi_{1'} \iota_{A'}  \notag \\
	&= \pi_{0'} (o_{A'} + (\frac{\pi_{1'}}{\pi_{0'}}) \iota_{A'}).
\end{align}
Moreover, the relations
\begin{align}
\frac{1}{\pi_{0'}} \mathcal{A}^{A'} \pi_{A'} &= \mathcal{A}^{0'} + \mathcal{A}^{1'} (\frac{\pi_{1'}}{\pi_{0'}}), \\
	\frac{1}{\pi_{0'}} \mathcal{B}^{A'} \pi_{A'} &= \mathcal{B}^{0'} + \mathcal{B}^{1'} (\frac{\pi_{1'}}{\pi_{0'}})
\end{align}
imply that equation (\ref{eqn:penroseTransform}) becomes an integral manifestly over $\mathbb{CP}^1$. Thus
\begin{align}
\varphi_{A'_1 \cdots A'_{2h}}(x) &= \frac{1}{2\pi i} \oint_\Gamma \frac { (o_{A_1}+(\frac{\pi_{1'}}{\pi_{0'}}) \iota_{A'_1}) \cdots (o_{A_{2h}}+(\frac{\pi_{1'}}{\pi_{0'}}) \iota_{A'_{2h}})} {(\mathcal{A}^{0'}+\mathcal{A}^{1'}(\frac{\pi_{1'}}{\pi_{0'}}))(\mathcal{B}^{0'}+\mathcal{B}^{1'}(\frac{\pi_{1'}}{\pi_{0'}}))^{2h+1} } d(\frac{\pi_{1'}}{\pi_{0'}}) \notag \\
	&= \frac{1}{2\pi i \mathcal{A}^{1'} (\mathcal{B}^{1'})^{2h+1}} \oint_\Gamma \frac{(o_{A_1}+\zeta\iota_{A'_1})\cdots(o_{A_{2h}}+\zeta\iota_{A'_{2h}})}{(\mu+\zeta)(\nu+\zeta)^{2h+1}} d\zeta
\end{align}
where $\zeta = \pi_{1'} / \pi_{0'}$, $\mu = \mathcal{A}^{0'} / \mathcal{A}^{1'}$, and $\nu = \mathcal{B}^{0'} / \mathcal{B}^{1'}$ represent the projective coordinate and poles respectively.

\begin{figure}[t] 
\centering
\includegraphics{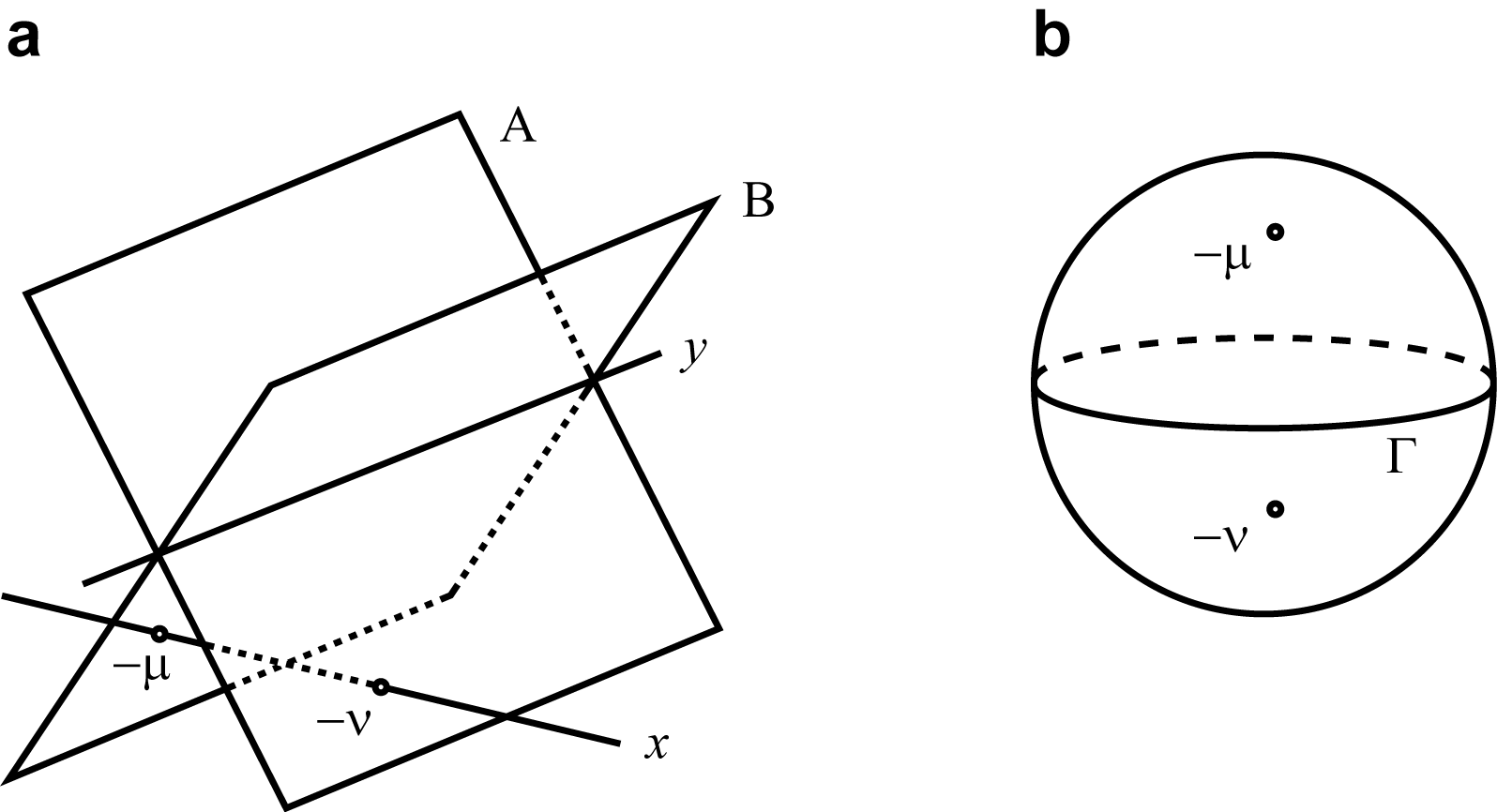}
\caption{Geometry of the integral in the Penrose transform. \textbf{a} In $\mathbb{PT}$, $A$ and $B$ are given as planes on which $f(Z)$ is singular, and they intersect at a $\mathbb{CP}^1$ describing a point $y \in \mathbb{CM}$. The point $x$ at which we are computing $\varphi$ is represented as a $\mathbb{CP}^1$ which meets $A$ and $B$ at the poles $-\mu$ and $-\nu$ of $f(Z)$. \textbf{b} The celestial sphere of $x$, where the contour $\Gamma$ separates the poles.}
\label{fig:integral}
\end{figure}

After the variable substitutions, the integral is straightforward. Taking $\Gamma$ to enclose $-\mu$ we have
\begin{align}
\varphi_{A'_1 \cdots A'_{2h}}(x) &= \frac{1}{\mathcal{A}^{1'} (\mathcal{B}^{1'})^{2h+1}} \underset{\zeta=-\mu}{\text{Res}} \frac{(o_{A_1}+\zeta\iota_{A'_1})\cdots(o_{A_{2h}}+\zeta\iota_{A'_{2h}})}{(\mu+\zeta)(\nu+\zeta)^{2h+1}} \notag \\
&= \frac{1}{\mathcal{A}^{1'} (\mathcal{B}^{1'})^{2h+1}} \frac{(o_{A'_1}-\mu\iota_{A'_1})\cdots(o_{A'_{2h}}-\mu\iota_{A'_{2h}})}{(\nu-\mu)^{2h+1}} \notag \\
&= \frac{(\mathcal{A}^{1'} o_{A'_1} - \mathcal{A}^{0'} \iota_{A'_1}) \cdots (\mathcal{A}^{1'} o_{A'_{2h}} - \mathcal{A}^{0'} \iota_{A'_{2h}})} {(\mathcal{A}^{1'}\mathcal{B}^{0'} - \mathcal{A}^{0'}\mathcal{B}^{1'})^{2h+1}} \notag \\
&= \frac{1}{(\epsilon_{A'B'}\mathcal{A}^{A'}\mathcal{B}^{B'})^{2h+1}}\mathcal{A}_{A'_1} \cdots \mathcal{A}_{A'_{2h}} \notag \\
&=  \left(\frac{2}{\mu_B \sigma^B (x^a-y^a)(x_a-y_a)}\right)^{2h+1} \mathcal{A}_{A'_1} \cdots \mathcal{A}_{A'_{2h}} \label{eqn:knotField}
\end{align}
where $ \mu_B \sigma^B$ is the constant $\Omega$ in the main text, and the point $y$ is given in a flagpole form as
\begin{equation}
y^{AA'} = i \frac{\sigma^A \lambda^{A'} - \mu^A \psi^{A'}}{\mu_B \sigma^B}.
\end{equation}
The contour integral, whose geometry is shown in Figure \ref{fig:integral}, thus turns the pole structure of $f(Z)$ at the point $x$ into the specific field configuration in equation (\ref{eqn:knotField}).

%
%

\end{document}